\documentclass[twocolumn,superscriptaddress,nofootinbib,aps,prl,floatfix,preprintnumbers,amsmath,amssymb,groupedaddress]{revtex4}

\usepackage{dsfont}
\usepackage{epsfig}
\usepackage{slashed}
\usepackage{bbold}
\usepackage{psfrag}
\usepackage{color}
\PassOptionsToPackage{caption=false}{subfig}
\usepackage{subfig}
\usepackage{multirow}
\usepackage{booktabs}

\begin{document}
\title{Light non-degenerate squarks at the LHC}

\author{Rakhi Mahbubani}
\affiliation{Theory Division, CERN, 1211 Geneva 23, Switzerland}
\author{Michele Papucci}
\affiliation{Theoretical Physics Group, Lawrence Berkeley National
  Laboratory, Berkeley CA 94720}
\affiliation{Department of Physics, University of California, Berkeley
CA 94720}
\author{Gilad Perez}
\affiliation{Theory Division, CERN, 1211 Geneva 23, Switzerland}
\affiliation{Department of Particle Physics and Astrophysics, Weizmann
Institute of Science, Rehovot 76100, Israel}
\author{Joshua T. Ruderman}
\affiliation{Theoretical Physics Group, Lawrence Berkeley National
  Laboratory, Berkeley CA 94720}
\affiliation{Department of Physics, University of California, Berkeley
  CA 94720}
\author{Andreas Weiler}
\affiliation{DESY, Notkestrasse 85, D-22607 Hamburg, Germany}

%\pacs{PACS}
%\keywords{keywords}
\preprint{CERN-PH-TH/2012-320, DESY 12-231}
\begin{abstract}
Experimental bounds on squarks of the first two generations assume their masses to be eightfold degenerate, and consequently constrain them to be
heavier than $\sim 1.4$ TeV when the gluino is lighter than $2.5$ TeV.
The assumption of squark-mass universality 
is neither a direct consequence of Minimal Flavor Violation (MFV), which allows for splittings within squark generations,
nor a prediction of supersymmetric alignment models, which allow for splittings between generations.  
We reinterpret a recent CMS multijet plus missing energy
search allowing for deviations from $U(2)$ universality, and find significantly weakened squark bounds: a $400 \,$GeV second-generation
squark singlet is allowed, even with exclusive decays to a massless
neutralino; and in an MFV scenario, the down-type squark singlets can be
as light as $600 \, $GeV provided the up-type singlets are pushed up to
$1.8 \, $TeV, for a $1.5$~TeV gluino and decoupled doublet squarks.
\end{abstract}
\maketitle
%%%%%%%%%%%%%%%%%%%%%%%%%%%%%%%%%%%%%%%%%%%%%%

{\it\bf Introduction:} As a solution to the electroweak gauge hierarchy problem in the Standard Model (SM), supersymmetry (SUSY) is an immensely compelling paradigm. However its most popular incarnation, the Minimal Supersymmetric Standard Model (MSSM), has a huge parameter space with over a hundred independent parameters.  A rather small subset of this parameter space is constrained by naturalness~\cite{Dine:1993np, Dimopoulos:1995mi, Cohen:1996vb}, and it is possible to reinterpret current searches in the context of an effective theory containing only the most relevant degrees of freedom~\cite{Papucci:2011wy,Brust:2011tb}.
The requirement of naturalness, however,  
gives no guidance regarding the vast majority of parameters of the MSSM or its extensions: 
those related to the squarks of the first two generations
are largely insignificant to the naturalness argument due to the smallness of their Yukawa couplings.  At face value, ATLAS and CMS simplified model searches disfavor first- and second-generation squarks below $\sim 1.4 \,$TeV, for a gluino mass of $2.5$ TeV or less.   However, these limits assume an eightfold degeneracy for the masses of two flavors  of electroweak doublets, $\tilde{Q}_{1,\,2}$, and up- and down-type singlets, $(\tilde{u}_R,\,\tilde{c}_R)$ and $(\tilde{d}_R,\,\tilde{s}_R)$.  We will argue below that this assumption is not justified.
%It is important to ask whether the assumption of eightfold degeneracy underpinning these bounds is justified.  We will argue below that it is not. 

One might wonder how drastic the practical consequences of relaxing
this assumption would be. A naive rescaling, assuming that the squark
production cross section goes like $m_{\tilde{q}}^{-6}$, would suggest
that the bound on a single squark degree of freedom should be around
$8^{-1/6}\sim 30\%$ smaller than the current limit, hardly a dramatic change.
This estimate fails to take into account two important effects: the first is the drop in signal efficiency at low squark mass due to the hard cuts necessary to minimize SM backgrounds; the second is due to Parton Distribution Functions (PDFs), which, for non-decoupled gluinos, result in the cross sections for second-generation squark production being smaller than the corresponding ones for the first generation (previously noted in the context of Dirac gluinos by~\cite{Heikinheimo:2011fk,Kribs:2012gx}). Both these effects work in concert to further weaken the bound on a second-generation squark with respect to the naive expectation, and make, in themselves, a strong case for reanalyzing the data in the context of non-degenerate light squarks.
\begin{figure*}[t!]
\centering
\includegraphics[width=0.9\linewidth]{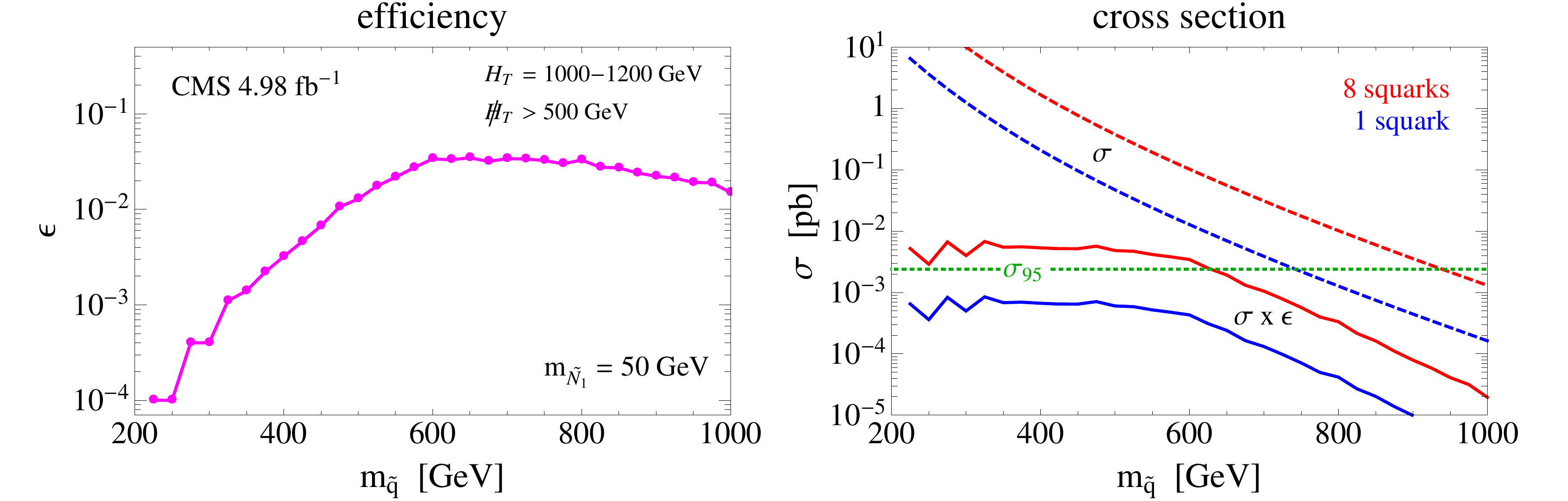}
\caption{(Left panel) Variation of signal
  efficiency$\times$acceptance ($\epsilon$) with squark mass for a single channel in the CMS multijet + MET search with 4.98 fb$^{-1}$ of data, for a squark 
  simplified model with a 50 GeV neutralino.  (Right panel) Cross section $\sigma$ (dotted lines), and $\sigma\times\epsilon$ for the chosen channel (solid lines) for 8 squark degrees of freedom (in red), naively rescaled for a single degree of freedom (in blue).  The fiducial cross section limit is $\sigma_{95}$ (dotted green).  The steeply falling efficiencies at low squark masses result in a significant reduction of the limit for the rescaled cross section.\label{fig:eff}}
\end{figure*}

Adding the theoretical perspective only makes the case more compelling. Squark-mass universality is motivated solely by high scale SUSY breaking models such as Minimal Supergravity, and is not required to solve the SUSY flavor problem. Indeed it is not a direct consequence of Minimal Flavor Violation (MFV) (see {\it e.g.}~\cite{D'Ambrosio:2002ex,Buras:2003jf}), which allows for splittings between squarks belonging to different representations of the SM gauge group, lifting the eightfold degeneracy to a $4+2+2$ pattern.
Furthermore, SUSY alignment models, which address the
SM flavor puzzle as well as the SUSY flavor problem,
naturally predict an anarchic sfermion spectrum, with
O(1) splittings between squarks. This is achieved by assigning particular $U(1)$ charges to the different squark
generations under a new set of flavor (`horizontal') symmetries~\cite{Nir:1993mx}, which forces the squark doublet soft masses to `align with' (be diagonal in) the down Yukawa mass basis, trivially satisfying the most severe constraints from CP violation in the Kaon and $B$ systems and relaxing the constraint on the mass splitting (see {\it e.g.}~\cite{Nir:2002ah}). The latter was previously thought to be strongly constrained at the percent level~\cite{Nir:2007xn,Blum:2009sk}, due to a combination of bounds from $K-\bar K$ and $D-\bar D$ mixing. However, it was recently understood that the bounds from CP-violating processes are largely avoided by successful alignment models~\cite{Gedalia:2012pi}, leading to a much weaker constraint on squark mass
non-degeneracies, for doublets as well as singlets.

To summarize, there is strong practical, as well as theoretical, motivation for re-examining the experimental bounds on the masses of the first two generations of squarks when the assumption of a full eightfold degeneracy is relaxed.  In the remainder of this paper we describe in detail our methods for reinterpreting current LHC searches within this more general framework, and present estimated limits for a number of distinct, phenomenologically interesting scenarios with non-degenerate squarks.  We end with a comparison of the effectiveness of the different light squark search strategies at ATLAS and CMS, and comment on the possibility of optimizing such searches by readjusting analysis cuts.

\begin{figure*}
\centering
\includegraphics[width=0.8\linewidth]{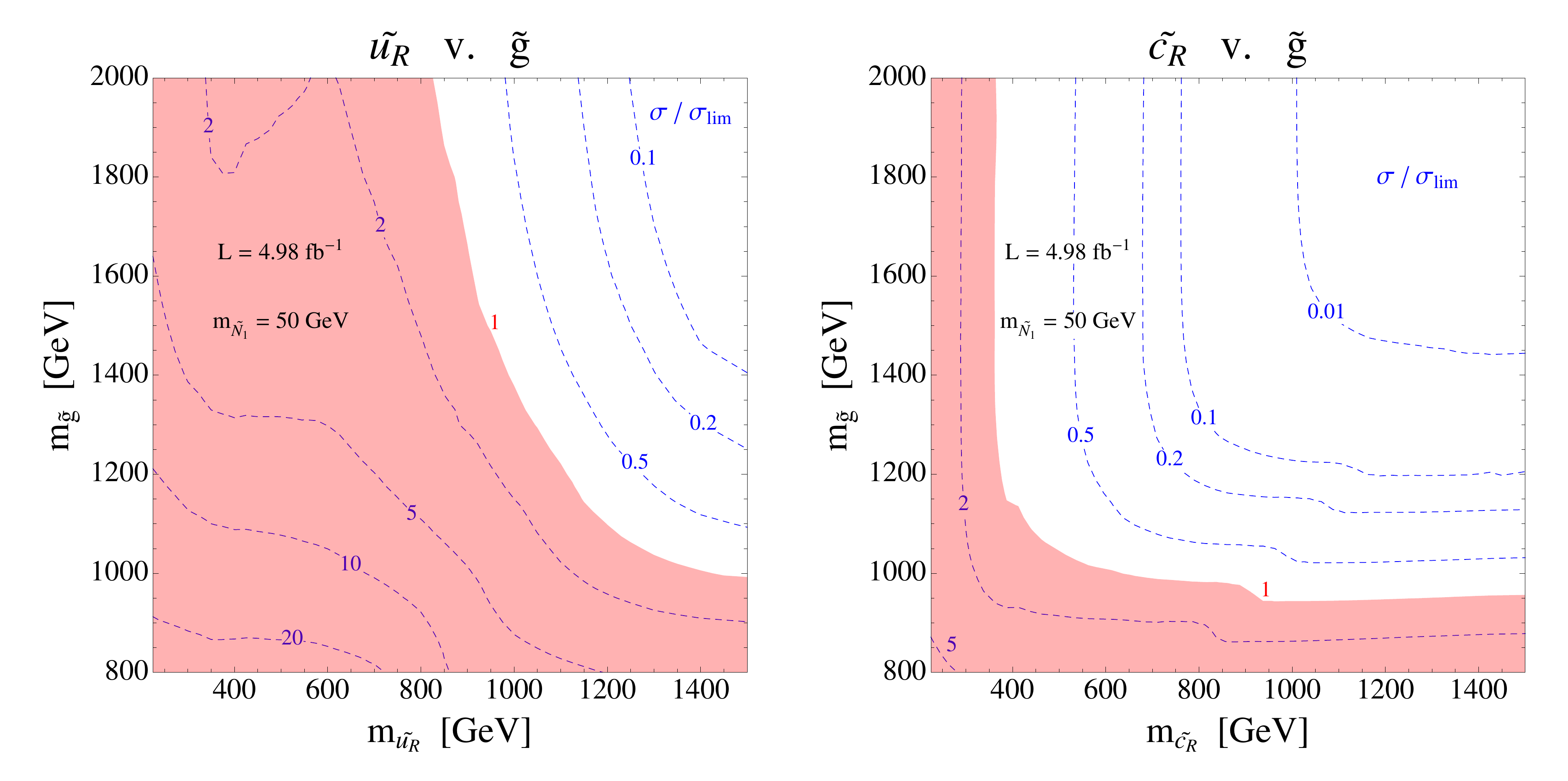}
\caption{The limits in the squark-gluino plane for production and decay of a single squark degree of freedom of different flavors.  The effect of the PDF enhancement on the production of a first-generation squark, via a $t$-channel gluino, is clearly manifest in the form of a more stringent limit for fixed gluino mass; this difference approaches zero in the gluino decoupling limit.\label{fig:PDF}}
\end{figure*}

{\it\bf Procedure:} We determine limits on simplified spectra consisting of a gluino and one or more squarks, decaying to a massless neutralino plus jets, with other superpartners decoupled. We focus on the recent CMS 7 TeV multijet plus missing energy (MET) search with 4.98 fb$^{-1}$ of integrated luminosity~\cite{:2012mfa}; we compare this with the performance of other CMS and ATLAS jets plus MET searches below. For the squark-neutralino and gluino-neutralino simplified models, CMS makes available efficiency maps in the $(m_{\tilde{q}},m_{\chi_0})$, $(m_{\tilde{g}},m_{\chi_0})$ planes: we use this information directly whenever possible. Our analysis also requires efficiency maps for a squark-gluino-neutralino simplified model, with on-shell intermediates in decays where relevant, as well as for mixed production of two squarks of different masses.  Lacking the pertinent efficiency and acceptance information in these cases, we simulate them using Pythia 6.4.24~\cite{Sjostrand:2006za} with the CTEQ6L1  PDF set~\cite{Pumplin:2002vw}, subsequently passing the events through two different analysis pipelines, ATOM (``Automatic Test of Models'')~\cite{Bauer:2012rr}, and the PGS~\cite{pgs} `theorist-level' detector simulation, as an internal cross-check.  The former (soon-to-be publicly available) is a RIVET~\cite{Buckley:2010ar}-based tool that estimates current LHC limits on a given model, flagging problematic regions {\it e.g.} where the signal leaks into control regions, or where signal efficiencies are too sensitive to cut positions.  We validate both pipelines by reproducing the published limits of the search within $\sim$ 50 GeV, corresponding to an accuracy in the estimated experimental acceptances of better than 20\%. We correct the various production cross sections using NLO+NLL K-factors from NLLfast~\cite{Beenakker:1996ch,Kulesza:2008jb,Beenakker:2009ha,Beenakker:2011fu} in all figures except Fig.~\ref{fig:PDF}, where we use Prospino 2.1~\cite{Beenakker:1996ch}.  For mixed production, which is not considered in NLLfast/Prospino, we cannot use such a procedure reliably. In this case we present the result using the LO cross section, and show the effect of including K-factors of $1.5$ and $2.0$. We have checked using MadGOLEM~\cite{GoncalvesNetto:2012yt} that, for a selection of points close to the current limit, the chosen K-factors are in fact conservative. 
Finally, at each point in signal parameter space, we combine the fourteen non-overlapping signal regions using an approximated likelihood $\mathcal{L} = \Pi_{i=1}^{14} {\rm poiss}(n^{obs}_i | s_i + \bar b_i) \cdot {\rm gauss}(\bar b_i |  b_i^{exp},\delta b_i)$. All limits are 95\% confidence-level exclusions derived from a profile likelihood ratio using the $CL_s$ technique~\cite{Read:2000et}, computed with a private toy MC code, verified by comparing it to asymptotic distributions~\cite{Cowan:2010js} calculated by RooStats~\cite{Moneta:2010pm}.  

Note that the NLO squark production cross section, as computed by Prospino 2.1 in the limit of large gluino mass, decouples much more slowly than one would expect ($\sigma \propto m_{\tilde g}^{-2} $), leading to an overestimate of the K-factors in this corner of parameter space.  When using Prospino, we are careful to stay in the safe region.
Note also that the naive use of Pythia, with all the squarks and gluino production channels enabled and non-degenerate squark spectra can result in under-sampling of certain regions of phase space. As a workaround, we performed multiple runs corresponding to different sub-processes.

{\it\bf Results:} As mentioned above, naively rescaling the squark cross section limit by the number of squark degrees of freedom in the first two generations significantly underestimates the change in the limit for a single squark for two non-trivial reasons.  The first is the sharp drop in experimental efficiencies 
at small squark masses.  This is due to hard cuts (mainly on variables correlated with the visible and invisible energy of the event, such as $M_{\rm eff},H_{T},E\!\!\!\!/_{T},H\!\!\!\!/_{T}$) placed on the data in order to suppress the large SM backgrounds.  We illustrate this point in Fig.~\ref{fig:eff}.  The left panel shows the variation of the CMS-provided efficiency $\times$ acceptance ($\epsilon$) with squark mass for a chosen channel in the squark simplified model analysis, with a neutralino mass of $50\,$GeV. The steep decrease in signal efficiency for squark masses below about 500 GeV seen in this plot is typical of all pertinent cut-and-count searches, making these particularly inefficient for light squarks.  The right panel shows the cross section $\sigma$ (dotted lines), and $\sigma\times\epsilon$ (solid lines) for the chosen channel for 8 squark degrees of freedom (in red), naively rescaled for a single degree of freedom (in blue), with decoupled gluinos.  The fiducial cross section limit in this region, $\sigma_{\rm 95}$, is indicated by the dotted green line.  The rising cross section with decreasing squark mass is compensated by the steeply falling efficiency, significantly reducing the exclusion limit
for the rescaled estimate.  Note that this is not the true bound on a single squark, but simply illustrative of the repercussions of the falling efficiencies.

The second important reason is due to PDFs:  the large valence quark density in the proton leads to a squark pair-production cross section that is dominated by first-generation squark production with gluinos in the $t$-channel, if the latter are accessible.  The current limit for non-decoupled gluinos is therefore driven predominantly by first-generation up-type squarks, with limits on the second-generation being correspondingly weaker. We neglect squark mixing for simplicity (see conclusion for comment).  The PDF effect is shown in Fig.~\ref{fig:PDF}, where we plot contours of $\sigma/\sigma_{\rm lim}$, the ratio of the total cross section to the excluded cross section, in the squark-gluino plane, for a $50\,$GeV neutralino, and a single squark degree of freedom of varying flavor.  The estimated limit is much weaker for a second-generation squark, where the cross section has no PDF enhancement.  This difference is expected to decrease with increasing gluino mass, as the contribution from the $t$-channel gluino decreases, but due to the slow decoupling of the gluino, the asymptotic behavior is reached for gluino masses larger than those shown here.

\begin{figure*}[t!]
\centering
\includegraphics[width=0.95\linewidth]{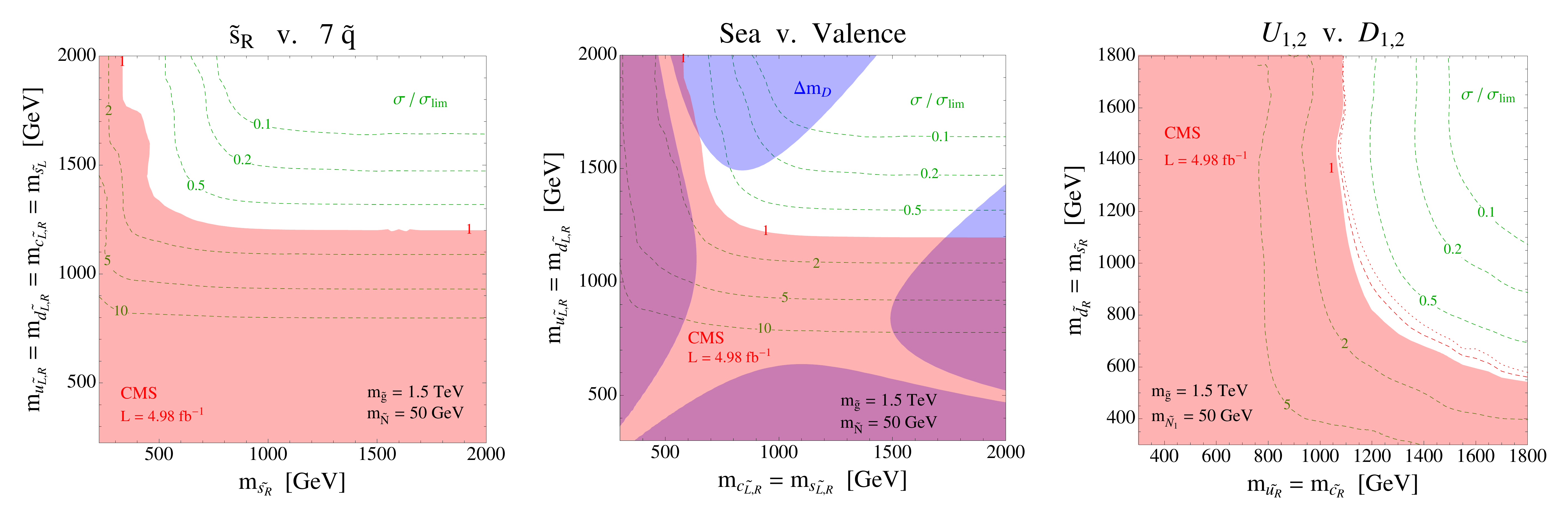}
\caption{Squark mass limits in three phenomenologically interesting scenarios with non-degenerate first- and second-generation squarks.  The left panel contains the least constrained scenario, with a single second-generation squark flavor split from all others; the middle panel corresponds to an alignment-type scenario with first-generation squarks split from the second-generation.  The shaded blue region is excluded by flavor and CP violation constraints which apply to electroweak doublet squarks only, while the singlet spectrum remains completely unconstrained; the right panel corresponds to an MFV-type scenario with split up-type and down-type singlets, and doublets formally decoupled. The red dashed (dotted) lines represent the exclusion contour if the LO mixed up-down squark production cross section is multiplied by a K-factor of $1.5$ ($2.0$).\label{fig:3cases}}
\end{figure*}
The striking contrast between the first- and second-generation squark bounds is due in part to the fact that we are taking an artificial limit in which all other squark degrees of freedom are formally decoupled.  For a more generic, anarchic squark spectrum, one would expect contributions to the total production cross section from all squark flavors.  In Fig.~\ref{fig:3cases}, we show the estimated limits in three more `realistic' and phenomenologically interesting scenarios with multiple non-degenerate squark degrees of freedom, and a gluino just above the current limit, at $1.5\,$TeV.  The left-hand panel covers the least constrained scenario of a single light second-generation squark split from all the other squark flavors, all of which are accessible. The middle panel gives the estimated limits in an alignment-type scenario with first-generation squarks split from the second generation.  It illustrates an interesting interplay between flavor and collider physics, since the splitting between the electroweak doublets cannot be arbitrarily large due to the combination of constraints from $K-\bar K$ and $D-\bar D$ mixing, assuming down alignment~\cite{Gedalia:2012pi}. Note that the flavor constraints shown in the plot include the full dependence on the squark masses, crucial when the splitting is large~\cite{nonlinear}.
Although the singlet squarks are kept degenerate with the corresponding doublets for simplicity, their splittings are unconstrained by flavor, and they could also be decoupled, resulting in weaker LHC bounds (corresponding to the contour $\sigma/\sigma_{\rm lim} \sim 2$), with unchanged flavor bounds.  The right-hand panel contains the limits in an MFV-type scenario, with split up-type and down-type singlets, and doublets formally decoupled. The red dashed (dotted) lines represent the exclusion contour if the LO mixed up-down squark production cross section is multiplied by a K-factor of $1.5$ ($2.0$).
\begin{figure}[h!]
\centering
\includegraphics[width=0.45\textwidth]{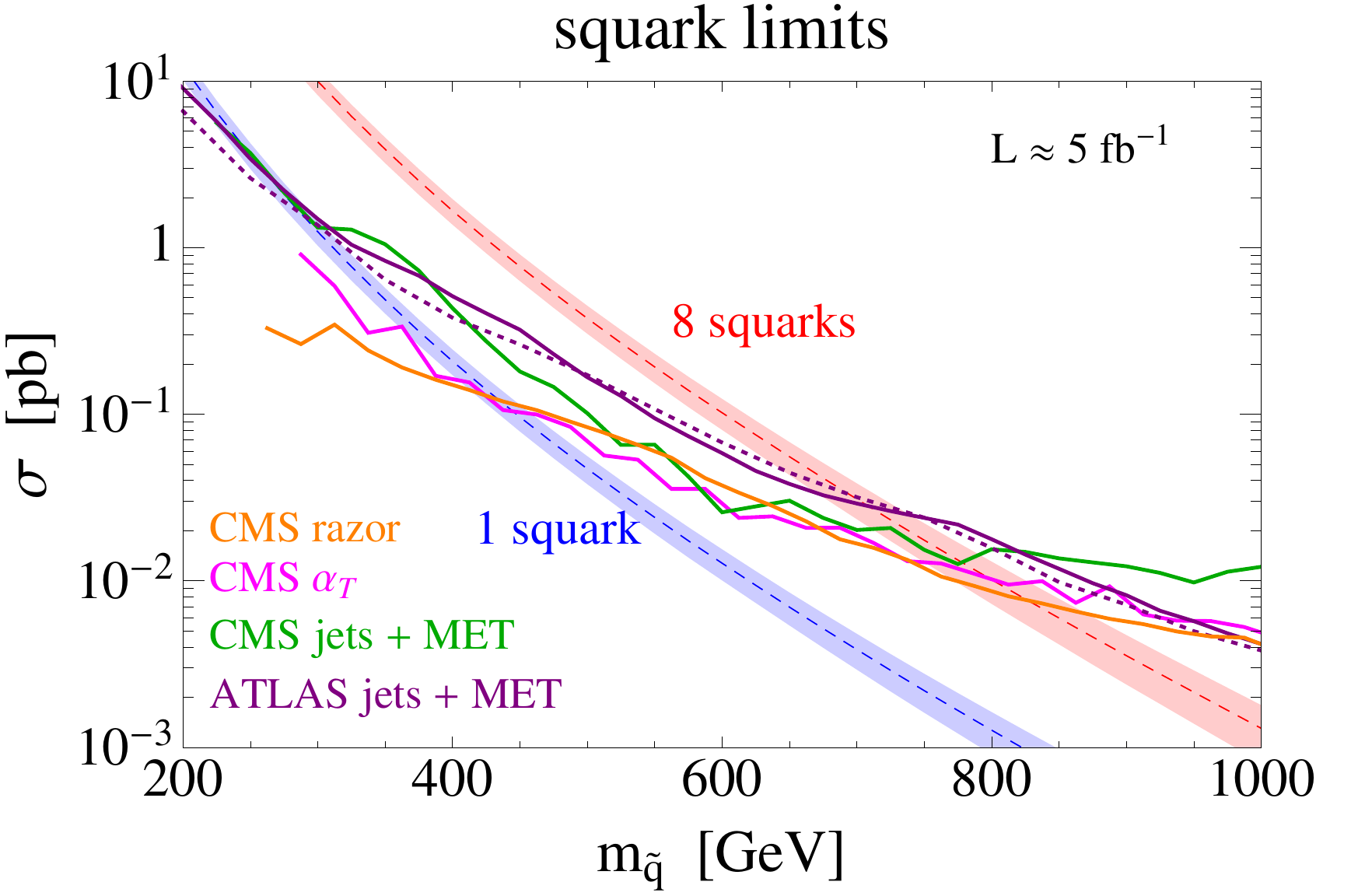}
\caption{Comparison between upper limits on squark pair-production cross sections with a decoupled gluino and massless neutralino, from $7\,{\rm TeV}$ $5\,{\rm fb}^{-1}$ ATLAS and CMS jets plus MET searches~\cite{:2012rz,CMS-PAS-SUS-12-005,:2012mfa,Chatrchyan:2012wa}.  We use the official experimental limits, except for the ATLAS search where we use our estimate of the limit, simulating the search with ATOM (solid) and PGS (dotted).
\label{fig:compare}
}
\end{figure}

The surprisingly weak limits, in particular for squarks of the second generation, demonstrate how ineffective current searches are for light squarks.  Re-optimizing the ATLAS 2-6 jets plus MET search using only the $m_{\rm eff}$ cut is not effective: while the background grows like $m_{\rm eff}^6$, the signal grows much more slowly, ensuring that decreasing the $m_{\rm eff}$ cut makes things worse.  It is possible that the limits would improve on performing either a full re-optimization including all cut variables, or a shape analysis; such a study, however, is beyond the scope of this paper.  Instead, in Fig.~\ref{fig:compare}, we compare the limits for squark cross sections from various 7 TeV ATLAS and CMS jets plus MET searches (which have limits for degenerate squarks that are competetive with those of recent 8 TeV searches~\cite{ATLASCONF5fb,CMSCONF11fb}).  We find indeed that the most stringent bounds come from the more complex shape-based analyses, such as the CMS razor search.

{\it\bf Conclusion:} We have argued that a combination of reduced efficiencies and suppression due to PDFs leads to constraints on non-degenerate squark masses (for the first two generations) that are significantly weaker than those assuming eightfold degeneracy.  For instance, an ${\cal O} (400{\rm\, GeV })$ squark belonging to the second generation can be buried in the LHC jets plus MET data. 
In the above analysis we have neglected for simplicity the effects of squark mixing, which could be sizable in alignment models. In addition, our reinterpreted limits, while assuming the bino is the lightest SUSY particle (LSP), are still applicable for singlino or gravitino LSPs, or when additional electroweak ({\it e.g.} higgsinos) and leptonic states are present, but do not drastically alter the light squark branching ratios.
In spite of the dramatic increase of the squark pair-production cross section with decreasing squark mass, the corresponding SM backgrounds tend to grow even faster, and dedicated searches will be required in order to extract light squarks from the data.  One possible addition to these is charm tagging, which would improve the sensitivity for light charm squarks. This is already implemented in Tevatron $\tilde t\rightarrow c \chi^{0}$ searches~\cite{Abazov:2008rc,Aaltonen:2012tq}, which set limits for a single charm squark which are weaker than those obtained here.

The impact of observing light squarks would go beyond `mere' discovery of SUSY (a dramatic event by itself), since the knowledge of a split two-generation spectrum encodes information about SUSY breaking.  Moreover, discovering a splitting between generations (breaking of the $U(2)$ flavor symmetry) would have sensational implications, giving us insight on microscopic flavor dynamics far beyond the direct reach of the LHC.

%\begin{acknowledgments}
{\it\bf Acknowledgments:} We thank C.~Autermann, M.~d'Onofrio, W.~Ehrenfeld, A.~Hoecker, P.~Pralavorio, S.~Sharma, and G.~Weiglein for discussions. We thank T.~Plehn and D.~Lopez-Val for providing comparisons with MadGOLEM, and A.~Kulesza and S.~Thewes for support with NLLfast. RM thanks J.~Alwall, J.~Gallicchio, E.~Izaguirre, O.~Mattelaer, V.~Sanz and P.~Skands for helpful conversations.  RM, MP and AW thank the CERN TH group for its warm hospitality. MP thanks the Aspen Center for Physics where some of these results were obtained. GP is the Shlomo and Michla Tomarin development chair, supported by the grants from GIF, Gruber foundation, IRG, ISF and Minerva.  JTR is supported by a fellowship from the Miller Institute for Basic Research in Science. The work of AW was supported in part by the German Science Foundation (DFG) under the Collaborative Research Center (SFB) 676. MP is supported in part by the Director, Office of Science, Office of High Energy Physics, of the US Department of Energy under Contract DE-AC02-05CH11231.
%\end{acknowledgements}
%%%%%%%%%%%%%%%%%%%%%%%%%%%%%%%%%%%%%%%%%%%%%%
\bibliography{nondeg_Lett}
%%%%%%%%%%%%%%%%%%%%%%%%%%%%%%%%%%%%%%%%%%%%%%
%%%%%%%%%%%%%%%%%%%%%%%%%%%%%%%%%%%%%%%%%%%%%%
\end{document}